\documentstyle[aps,multicol,psfig,prb]{revtex}
\begin{document}

\draft

\noindent{To Mrs./Miss Aitken}    

\noindent{Editorial Office of Solid State Communications}  

\noindent{Oxford, GB}

\vskip 0.5truecm

\noindent{Dear Mrs./Miss Aitken,}          

here enclosed please find the manuscript 0175 cc
\vskip 0.3truecm

``Optical conductivity of CuO$_2$ infinite-layer films''
\vskip 0.3truecm

by P. Dore {\it et al.},

that I send you again, for your convenience, by also including the figures in 
epsf format. In the previous submission, I sent you the latex file only, while
the figures were sent by fax.

\vskip 0.5truecm
\hskip 8.5truecm                          Sincerely yours,

\vskip 0.5truecm

\hskip 8.5truecm                          Prof. P. Calvani
\vskip 0.5truecm

Please send all correspondence concerning this paper to:

Prof. P. Calvani, 

Dipartimento di Fisica,  Universita' ``La Sapienza''

2, Piazzale A. Moro

00185 Roma, Italy  
                 
\vskip 0.2truecm 
(Fax n. +39-6-4463158 - e-mail: P.CALVANI@CASPUR.IT)

\vfill

\newpage

\title{Optical conductivity of CuO$_2$ infinite-layer films}
\author{P. Dore, P. Calvani, G. De Marzi, A. Paolone\cite{Paolone},
and S. Lupi} 
\address{Istituto Nazionale di Fisica della Materia and} 
\address{Dipartimento di Fisica, Universit\'a di Roma ``La Sapienza,''
Piazzale A. Moro 2, 00185 Roma, Italy}
\author{G. Balestrino, G. Petrocelli, and R. Desfeux\cite{Desfeux}} 
\address{Istituto Nazionale di Fisica della Materia and  
Dipartimento di Scienze e Tecnologie Fisiche ed Energetiche,  
Universit\'a di Roma ``Tor Vergata'', Via della Ricerca Scientifica, 00133 
Roma, Italy}

\maketitle

\begin{abstract}
The infrared conductivity of CaCuO$_2$, SrCuO$_{2-y}$, and 
Sr$_{0.85}$Nd$_{0.15}$CuO$_2$ infinite-layer films is obtained from
reflectivity measurements by taking into account the substrate
contribution. SrCuO$_{2-y}$ and Sr$_{0.85}$Nd$_{0.15}$CuO$_2$
exhibit extra-phonon modes and structured bands in the midinfrared, not found 
in stoichiometric CaCuO$_2$. These features mirror those observed in the 
perovskitic cuprates, thus showin that the polaronic 
properties of High-$T_c$ superconductors are intrinsic to the CuO$_2$ planes.


\end{abstract}

\begin{multicols}{2}

Since the early observations by photoinduced absorption,\cite{Kim,Taliani} 
it has been pointed out that the injection of charges into the parent compounds
of High-$T_c$ cuprates produces unexpectd features in the infrared 
absorption of the $a-b$ planes. These features, similar to those previously
observed in other polar materials like AgBr and attributed to 
polarons,\cite{Devreese} include extra-phonon lines (PILM's, PhotoInduced 
Local Modes), and a broad band centered between 1000 and 1500 cm$^{-1}$. 
Afterwards, Infrared Active Vibrations (IRAV) similar to the PILM's, together
with bands peaked around 1300 cm$^{-1}$ and named $d$ or $J$, have been 
observed in a variety of chemically-doped 
cuprates.\cite{Thomas92,Falck,Bucher,prb96,Salje} Both kinds of bands 
have been shown to survive in their metallic phase.\cite{Euro} The IRAV, or 
PILM, peaks have been attributed to local vibrations of clusters of atoms 
distorted by the injected charges, which self-trap under a strong coupling with
the polar lattice.\cite{Yonemitsu} The $d$ band at higher energies is then 
straightforwardly explained in terms of photoexcited hopping of those charges, 
in agreement with the polaronic models of the optical 
conductivity\cite{Eagles,Emin,Alexandrov} and with the observation of a 
fine structure at low doping.\cite{prb96}    

In connection with the present debate on the role of polarons in 
High-$T_c$ superconductivity, it may be crucial to determine 
whether the above infrared bands are originated in the CuO$_2$ 
planes, where superconductivity takes place, or in the surrounding crystal 
structures (often pictorially indicated as the ``out-of-plane stuff''). 
According to some authors, the polaronic transport in High-$T_c$ cuprates 
should only involve the Cu-O chains, where present.\cite{Ranninger}  
This point of view is supported, for instance, by the  absorption spectra of 
Y$_2$Ba$_4$Cu$_{6+n}$O$_{14+n}$, where a strong $d$ band is observed when the 
radiation is polarized along the chains.\cite{Bucher}  In other systems like 
La$_{2-x}$Sr$_x$CuO$_{4+y}$ and Bi$_2$Sr$_2$CaCu$_2$O$_8$,\cite{Bianconi} 
small polarons with heavy masses are invoked to describe the observed periodic 
displacements of the apical oxygens relative to the Cu-O planes. 
On the other hand, the infrared observation of local modes and/or polaron-like
$d$ bands in compounds so different as 
Y$_2$Ba$_4$Cu$_{6+n}$O$_{14+n}$,\cite{Bucher} 
Bi$_2$Sr$_2$YCu$_2$O$_8$,\cite{prb96} La$_2$CuO$_{4+y}$,\cite{Falck} and  
Nd$_{2-x}$Ce$_x$CuO$_{4-y}$,\cite{Thomas92,prb96} suggests
that the polaronic phenomena involve directly the CuO$_2$ planes, independently
of the presence of chains or apical oxygens. 

In the present work we address this problem by measuring the optical 
conductivity of the simplest crystal structures which contain CuO$_2$ planes, 
the infinite-layer (IL) films of general formula (Ca,Sr)$_{1-x}$Nd$_x$CuO$_2$. 
A stoichiometric IL film consists of an infinite stack of CuO$_2$ planes 
separated by layers of alkaline-earth ions. The samples here prepared and 
measured are doped by reduction or by substitutions at the sites of the 
alcaline-earth ions, then no out-of-plane structures are present. Any 
doping-induced spectral feature can be originated only in the CuO$_2$ planes. 

The infinite-layer (IL) structure (Ca,Sr)CuO$_2$  is thermodinamically unstable
for most compositions under the usual solid state reactions. However, the 
corresponding epitaxial films have been successfully grown by Pulse Laser 
Deposition (PLD), Reactive Sputtering or Molecular Beam 
Epitaxy.\cite{Norton,Sugii,Liu} 
Recently, good quality CaCuO$_2$ films were grown directly on NdGaO$_3$, (110)  
oriented, substrates.\cite{Bale} For the present experiment, three IL films 
were grown by PLD on SrTiO$_3$, a substrate more suitable for optical 
measurements\cite{Donato} than NdGaO$_3$. Details on the deposition procedure 
are reported elsewhere.\cite{Bale} The first sample was a CaCuO$_2$ film 
to be identified in the following as CCO. Since such a film cannot be
deposited on bare SrTiO$_3$, we grew a SrTiO$_3$-SrCuO$_2$-CaCuO$_2$ 
heterostructure.\cite{Gupta,Aruta} The SrCuO$_2$ 
buffer layer, less than 5 nm thick, did not affect the infrared
measurements.  The CCO film had a thickness $d_f$$\simeq$ 100 nm and was grown  
under an O$_2$ partial pressure $p_O$ = 0.35 mbar.  
Then, a film of nominal composition SrCuO$_2$ (SCO) with $d_f \sim$ 200 nm 
was directly grown on SrTiO$_3$, under $p_O$ = 0.4 mbar. In SCO films, an 
oxygen mobility much higher than in CCO may cause either the inclusion of 
excess oxygen ions, or the formation of oxygen vacancies in the CuO$_2$ planes 
during the growth or the cooling down of the film.\cite{Gupta} In the present 
case, the oxygen pressure was kept very low during both phases of the film
preparation, in order to obtain oxygen deficient CuO$_2$ planes,\cite{Gupta}
i. e. a light electron doping. The resulting resistivity was 
$\rho \sim$ 10 $\Omega$ cm at room temperature. In the following, the chemical 
formula of this film will then be written as SrCuO$_{2-y}$. 
Finally, a Sr$_{0.85}$Nd$_{0.15}$CuO$_2$ (SNCO) 
film with $d_f\simeq$ 500 nm was directly grown on SrTiO$_3$ under $p_O$ =
2x10$^{-3}$ mbar. Here, the Nd donors inject a large number of electrons in the
CuO$_2$ plane, so that the room temperature resistivity falls to
$\rho \sim$ 10 m$\Omega$ cm. The structural properties of the three samples 
were studied by use of a $\theta$-2$\theta$ diffractometer, by Scanning 
Electron Microscopy and by Energy Dispersive Spectroscopy (EDS). As shown in 
Fig. 1, all films have the $c$ axis perpendicular to the surface and are 
single phase. A tiny amount of a 2$\sqrt{2}a \times 2\sqrt{2}a$ 
superstructure\cite{Yakabe} was detected in the SNCO film, whose nominal 
composition was verified both by EDS and by measurements of the $c$ lattice 
parameter.\cite{Yakabe}

\begin{figure}
{\hbox{\psfig{figure=ssc97-1.epsf,width=8.5cm}}}
{\small Fig. 1. X-ray diffraction spectra of the three IL films here studied. 
The (001) and (002) peaks of the SrTiO$_3$ substrate are also indicated. 
A spurious peak from a tiny amount of the 2$\sqrt{2}a \times 2\sqrt{2}a$ 
superstructure in the SNCO film is marked by a star.}  
\end{figure}

The reflectivity measurements were performed in the frequency range 25-16000 
cm$^{-1}$ for CCO, in the range 25-12000 cm$^{-1}$ for SCO and SNCO. 
The electric field of the radiation was polarized in the plane of the film. 

\begin{figure}
{\hbox{\psfig{figure=ssc97-2.epsf,width=8.5cm}}}
{\small Fig. 2. The reflectivity spectra at 300 K of the three IL films here 
studied (dots) and the fitting curves which allow to extract the bare-film
optical conductivities (solid lines). The reflectivity of the bare SrTiO$_3$ 
substrate is reported for comparison by dashed lines.} 
\end{figure}

The room-temperature reflectance $R(\omega)$ of the three samples is plotted by 
dots in Fig. 2. For sake of comparison, the reflectance 
of the bare SrTiO$_3$ substrate\cite{Dore1} is also reported by dashed lines.
The SrTiO$_3$ contribution dominates the farinfrared (FIR) reflectance of all 
samples, as the penetration depth of  
the incident radiation is much larger than d$_f$. However, several features 
can be attributed to the superimposed IL films. The CCO reflectance (top) 
shows a well defined charge-transfer band above 10000 cm$^{-1}$, while  
the MIR reflectivity follows that of the substrate. In the FIR region, 
a few absorptions appear around 230, 345, and 575 cm$^{-1}$. In the   
SCO reflectivity (middle), the charge-transfer band is reduced in intensity, 
while new absorption features arise in the MIR region. Moreover, further 
deviations from the substrate reflectance become evident in the FIR region.  
Finally, in the strongly doped SNCO, $R(\omega$) is much different from that 
of the bare substrate at all frequencies. The CT band disappears, the 
absorption in the midinfrared grows considerably, while  
features similar to those of SCO are observed in the farinfrared. 

The procedure described and tested in Ref. \onlinecite{Agostinelli} 
has been followed to extract from the spectra of Fig. 2 the optical 
conductivity $\sigma$($\omega$) of the bare IL films. Shortly,  
the unknown dielectric function $\tilde \epsilon_{IL} (\omega)$ 
has been modeled through a standard Drude-Lorentz expansion, while the 
dielectric function\cite{Dore1,Kamaras} of SrTiO$_3$ 
$\tilde \epsilon_{sub} (\omega)$ has been extracted from reflectance and 
transmittance measurements.\cite{Dore2} The reflectances of the 
film-plus-substrate systems in Fig. 2 have then been fitted as described 
in Ref. \onlinecite{Agostinelli}, and the Drude-Lorentz parameters of 
$\tilde \epsilon_{IL} (\omega)$ are thus determined. In CCO and SCO 
$\tilde \epsilon_{IL} (\omega)$ is made up of Lorentzians peaked
at $\omega >0$, in SNCO it also includes a Drude term at $\omega = 0$. As 
shown in Fig. 2, the fits (solid lines) accurately reproduce the reflectance 
data of all samples.

\begin{figure}
{\hbox{\psfig{figure=ssc97-3.epsf,width=8.5cm}}}
{\small Fig. 3. The room-temperature optical conductivity of the 
CaCuO$_2$ infinite-layer film is reported by a solid line, and compared with 
the farinfrared $\sigma$($\omega$) of a Ca$_{0.86}$Sr$_{0.14}$CuO$_2$ single
crystal (dashed line), as extracted from the reflectivity data of 
Ref. \onlinecite{Tajima}. $E$, $B$, and $S$ label the external mode, the 
CuO$_2$ bending, and the CuO$_2$ stretching, respectively.}
\end{figure}

The real part $\sigma$($\omega$) of the optical conductivity, as obtained from  
$\tilde \epsilon_{IL} (\omega)$, is reported in Fig. 3 (solid line) for 
the CCO sample. It exhibits three phonon lines at 230, 345 and 573 cm$^{-1}$ 
corresponding to the TO external mode ($E$), to the bending ($B$) and 
stretching ($S$) mode of the CuO$_2$ planes, respectively. The symmetric 
phonon lineshapes and the sharp charge-transfer band at 10500 cm$^{-1}$ 
confirm that this CCO film is stoichiometric, even if
a weak midinfrared absorption in the 1500 cm$^{-1}$ region may be attributed to 
a tiny amount of oxygen defects (see below). No infrared data on pure CaCuO$_2$
crystals are available to check our procedure, due to the above mentioned
instability of this compound. In Fig. 3, the farinfrared $\sigma$($\omega$)
extracted from the reflectivity\cite{Tajima} of a single crystal of 
Ca$_{0.86}$Sr$_{0.14}$CuO$_2$ is then reported for comparison.  Therein, the 
three TO phonon peaks are centered at 230 ($E$), 360 ($B$), and 583 cm$^{-1}$ 
($S$). The agreement between the spectrum of the single crystal and that of 
the IL film confirms the reliability of the present analysis. The slight 
shifts of the $B$ and $S$ phonon peaks of Ca$_{0.86}$Sr$_{0.14}$CuO$_2$ 
with respect to the corresponding peaks of CaCuO$_2$ can be attributed to an 
average expansion of the CuO$_2$ lattice in the former compound, in presence 
of the large Sr impurities.\cite{Tajima} 

\begin{figure}
{\hbox{\psfig{figure=ssc97-4.epsf,width=8.5cm}}}
{\small Fig. 4. The room-temperature optical conductivities of the IL films 
SrCuO$_{2-y}$ (top) and Sr$_{0.85}$Nd$_{0.15}$CuO$_2$ (center) are compared 
with that of a single crystal of Nd$_2$CuO$_{4-y}$ at 20 K (bottom, from Ref. 
\onlinecite{prb96}).  $E$, $B$, and $S$ label the external mode, the 
CuO$_2$ bending, and the CuO$_2$ stretching, respectively. The dots indicate
the local modes $\nu_1$ and $\nu_2$. The arrows mark the features resolved in
the spectrum of the slightly-doped SCO, and assigned to overtones and 
combinations of $\nu_1$ and $\nu_2$ (see text).}
\end{figure}
 
On the top of Fig. 4, $\sigma$($\omega$) is reported between 200 and 3000
cm$^{-1}$ for the slightly electron-doped film SrCuO$_{2-y}$. Major
changes are evident with respect to the stoichiometric CCO of Fig. 3. 
The bending mode transfers spectral weight to a softer vibration centered at 
$\nu_1$ = 305 cm$^{-1}$, while the stretching mode at 573 cm$^{-1}$ is nearly 
replaced by a strong peak at $\nu_2$ = 480 cm$^{-1}$. The two IRAV's 
(indicated by dots in the Figure) are not related to the replacement of 
the out-of-plane Ca by Sr, as confirmed 
by the phonon spectrum of Ca$_{0.86}$Sr$_{0.14}$CuO$_2$ in Fig. 3. On the
other hand, IRAV's are predicted to appear on the low-energy side of the 
corresponding extended phonons (on both sides for very strong 
coupling),\cite{Yonemitsu} whenever self-trapped charges are distorting the 
CuO$_2$ planes. The polaronic character of the SCO spectrum is confirmed 
by the insurgence above 1000 cm$^{-1}$ of a $d$ band with resolved phonon-like
structures. This effect has been theoretically 
predicted\cite{Devreese,Alexandrov} and already observed in perovskitic 
cuprates at low doping.\cite{prb96} Therein, however, the line spacing within 
the polaron band is narrower than in the IL film of Fig. 4, consistently with 
the higher number of vibrational modes available in the perovskite structure. 
The band structure should be based on
the local modes of the distorted clusters, not on the extended
phonons of the unperturbed lattice. Indeed, the frequencies of the features 
indicated by arrows in Fig. 4, and obtained by a fit to a sum of Lorentzian, 
correspond to the following 
series of overtones and combinations of the two IRAV's $\nu_1$ , $\nu_2$: 
2$\nu_1$ (observed at 635 vs. 610 cm$^{-1}$), 2$\nu_1$+$\nu_2$ (1100 
vs. 1090 cm$^{-1}$), 3$\nu_2$ (1450 vs. 1440 cm$^{-1}$), 2$\nu_1$ + 2$\nu_2$ 
(1630 vs. 1570 cm$^{-1}$), 4$\nu_2$ (1950 vs. 1920 cm$^{-1}$), 5$\nu_2$ 
(2350 vs. 2400 cm$^{-1}$). The excellent agreement here reported  
also implies that the eventual anharmonic shifts are not much larger than the 
experimental uncertainties. One may also notice 
that the combination $\nu_1$+$\nu_2$ and the overtone 2$\nu_2$,
expected around 785 and 960 cm$^{-1}$ respectively, are missing. 
Indeed, both these lines should appear in the region where the 
reflectivity of SrTiO$_3$ rapidly drops, and the sensitivity of our
procedure for reconstructing $\sigma (\omega)$ is poor.

In the heavily doped film Sr$_{0.85}$Nd$_{0.15}$CuO$_2$ (middle of Fig. 4),
the $E$ phonon is shielded by a broad Drude absorption with $\sigma (0)$ = 130 
$\Omega^{-1}$cm$^{-1}$ (in good agreement with the above dc determination 
of $\rho$) and with $\Gamma$ = 500 cm$^{-1}$. The $B$ and $S$ modes are barely
discernible, while the two IRAV's and the $d$ band are more intense than in
SCO. The $d$ band is also less resolved, but shows the peak $\nu_1$+$\nu_2$ 
at 800 cm$^{-1}$. This strong combination band, not seen in the SCO 
spectrum, is observed in heavily doped SNCO, probably due
to the smaller penetration depth of the radiation.
The overall $\sigma$($\omega$) of SNCO is 
impressively similar to those previously reported for the perovskitic  
cuprates.\cite{Thomas92,Falck,Bucher,prb96,Salje,Thomas93}
As an example, in the bottom of Fig. 4 we plot $\sigma$($\omega$) 
for an electron-doped single crystal of Nd$_2$CuO$_{4-y}$ (NCO) at 20 
K.\cite{prb96} The local modes of NCO associated with the modes $B$ and $S$ of 
its CuO$_2$ planes correspond to the IRAV's of the SNCO IL film. The
behaviors with frequency of their $d$ bands are also very similar. On the other
hand, the evolution with temperature of those bands seems to be different.
Both in Nd$_2$CuO$_{4-y}$ and La$_2$CuO$_{4+y}$, the IRAV's and the $d$ band 
are barely detectable at room temperature,\cite{prb96,Thomas93} and reach their 
maximum intensity at a ``polaron freezing'' temperature $T_0 \sim $ 150 K. 
In the SCO and SNCO infinite-layer films here investigated, preliminary  
measurements show that the optical conductivity does not change appreciably when
lowering the temperature, suggesting a $T_0 \geq $ 300 K. 
                                      
In conclusion, the present work first reports on the optical conductivity of 
CuO$_2$ infinite-layer films. In a CaCuO$_2$ film used as stoichiometric 
reference, $\sigma$($\omega$) just shows the three TO phonons and a sharp 
charge-transfer band in the near infrared. Two SrCuO$_2$ films have been 
electron-doped, one (SCO) by a few oxygen vacancies, the other (SNCO) 
by substitution of 
15\% Sr ions by Nd ions. In both cases the extended phonons and the 
charge-transfer band loose much of their intensity, while new features appear. 
These consist of a pair of IRAV modes in the farinfrared, and of a broad band 
above 1000 cm$^{-1}$. In the slightly-doped film, the latter has been resolved
in phonon-like peaks whose energies correspond to overtones and combinations
of the two IRAV's. In the strongly doped film, $\sigma$($\omega$) reproduces
faithfully that of electron-doped Nd$_2$CuO$_{4-y}$, a cuprate 
which contains out-of-plane oxygens. Remarkable correspondences can also be 
found with the spectra reported in the literature for hole-doped compounds with 
apical oxygens and chains. 

The similarities here reported between the  
spectra of infinite-layer compounds and those of different 
perovskitic cuprates provide conclusive evidence that the polaronic 
excitations observed in High-$T_c$ superconductors do 
not depend on the existence of chains, apical oxygens or other ``out-of-plane 
stuff'', but are intrinsic to the CuO$_2$ planes. This result removes a common 
objection to the possibility that polarons are directly involved 
in the mechanism of High-T$_c$ superconductivity.  

\acknowledgments
We wish to thank M. Capizzi and D. Emin for many useful discussions, and P.
Maselli for collaboration in the analysis of data. 
This work has been partially supported by the EU Human Capital 
\& Mobility Networks ``Development of Infrared synchrotron radiation''
(CHRXCT940551) and ``Chemical synthesis of novel superconductors'' 
(CHRXCT940461), as well as by the Advanced Research 
Programme HTSS of Istituto Nazionale di Fisica della Materia of Italy.


\end{multicols}


\begin{references}
\bibitem[*]{Paolone}Present address: Laboratoire pour l'Utilization du 
   Rayonnement Electromagn\'etique, Universit\'e Paris-Sud, 91405 Orsay Cedex, 
   France.
\bibitem[**]{Desfeux}Present address: Laboratoire Crismat - I. S. M. R. A., 6,
   Blvd. du Mar\'echal Juin, 14050 Caen Cedex, France
\bibitem{Kim} 
   Y. H. Kim, S-W. Cheong, and Z. Fisk, Phys. Rev. Lett. {\bf 67}, 2227 (1991);
   Y. H. Kim, A. J. Heeger, L. Acedo, G. Stucky, and F. Wudl, Phys. Rev. 
   B {\bf 36}, 7252 (1987).
\bibitem{Taliani}
   C. Taliani, R. Zamboni, G. Ruani, F. C. Matacotta, and K. I. Pokhodyna, Solid
   State Commun. {\bf 66}, 487 (1988).
\bibitem{Devreese}
   J. Devreese, in {\it Optical Properties of Solids}, ed. by E. D.
   Haidemenakis, Gordon and Breach, New York 1969, p. 16, and references 
   therein.
\bibitem{Thomas92}
   G. A. Thomas, D. H. Rapkine, S. L. Cooper, S-W. Cheong, A. S. Cooper,
   L. F. Schneemeyer, and J. V. Waszczak, Phys. Rev. B {\bf 45}, 2474 (1992).
\bibitem{Falck} 
   J. P. Falck, A. Levy, M. A. Kastner, and R. J. Birgenau, Phys. Rev. B 
   {\bf 48}, 4043 (1993).
\bibitem{Bucher}
   B. Bucher, J. Karpinski, E. Kaldis, and P. Wachter, Phys. Rev. B {\bf 45},
   000 (1992).
\bibitem{prb96}
   P. Calvani, M. Capizzi, S. Lupi, P. Maselli, A. Paolone, and P. Roy, Phys. 
   Rev. B {\bf 53}, 2756 (1996), and references therein.   
\bibitem{Salje}
   Y. Yagil and E. K. H. Salje, Physica C {\bf 256}, 205 (1996).
\bibitem{Euro}
   P. Calvani, M. Capizzi, S. Lupi, and G. Balestrino, Europhys. Lett. 
   {\bf 31}, 473 (1995).
\bibitem{Yonemitsu}
   K. Yonemitsu, A. R. Bishop, and J. Lorenzana, Phys. Rev. Lett. {\bf 69},
   965 (1992); Phys. Rev. B {\bf 47}, 8065 (1993).
\bibitem{Eagles}
   D. M. Eagles, Phys. Rev. {\bf 130}, 1381 (1963).
\bibitem{Emin}
   D. Emin, Adv. Phys. {\bf 24}, 305 (1975); Phys. Rev. B {\bf 48}, 13691 
   (1993).
\bibitem{Alexandrov}
   A. S. Alexandrov, V. V. Kabanov, and D. K. Ray, Physica C {\bf 224}, 247
   (1994).
\bibitem{Ranninger}
   J. Ranninger, J. M. Robin, and M. Eschrig, Phys. Rev. Lett. {\bf 74}, 4027
   (1995), and references therein.
\bibitem{Bianconi}
   A. Bianconi {\it et al.}, Phys. Rev. Lett. {\bf 76}, 3412 (1996), and
   references therein.   
\bibitem{Norton}
   D. P. Norton, B. C. Chakoumakos, J. D. Budai, and D. H. Lowndes, Appl. 
   Phys. Lett. {\bf 62}, 1679 (1993).
\bibitem{Sugii}
   N. Sugii, M. Ichikawa, K. Kubo, T. Sakurai, K. Yamamoto, and H. Yamauchi,
   Physica C {\bf 196}, 129 (1992).
\bibitem{Liu}
   Z. Liu, T. Hanada, R. Sekine, M. Kawai, and H. Koinuma, Appl. Phys. Lett. 
   {\bf 65}, 1717 (1994)
\bibitem{Bale}
   G. Balestrino, R. Desfeux, S. Martellucci, A. Paoletti, G. Petrocelli, A.  
   Tebano, B. Mercey, and M. Hervieu, J. Mater. Chem. {\bf 5}, 1879 (1995).
\bibitem{Donato}
   P. Calvani, M. Capizzi, F. Donato, P. Dore, S. Lupi, P. Maselli, and C.P.
   Varsamis, Physica C {\bf 181}, 289 (1991).
\bibitem{Gupta}
   A. Gupta, B. W. Hussey, T. M. Shaw, A. M. Gulay, M. Y. Chern, R. F. Saraf,
   and B. A. Scott, J. Solid State Chem. {\bf 112}, 113 (1994).
\bibitem{Aruta}
   C. Aruta, G. Balestrino, R. Desfeux, S. Martellucci, A. Paoletti, and G. 
   Petrocelli, App. Phys. Lett. {\bf 68}, 926 (1996). 
\bibitem{Yakabe}
   H. Yakabe, A. Kume, J. G. Wen, M. Kosuge, Y. Shionara, and N. Koshizuka, 
   Physica C {\bf 232}, 371 (1994).
\bibitem{Dore1}
   P. Dore, A. Paolone, and R. Trippetti, J. Appl. Phys. {\bf 80}, 000 (1996).
\bibitem{Agostinelli}
   P. Calvani, M. Capizzi, S. Lupi, P. Maselli, and E. Agostinelli, Physica 
   C {\bf 180}, 116 (1991).
\bibitem{Kamaras}
   K. Kamaras, K. L. Barth, F. Keilmann, R. Henn, M. Reedyk, C. Thomsen, M.
   Cardona, J. Kircher, P. L. Richards, and J. L. Stehle, J. Appl. Phys.
   {\bf 78}, 1235 (1995).
\bibitem{Dore2}
   P. Dore, G. De Marzi, and A. Paolone, Int. J. IR \& MM Waves, in press.
\bibitem{Tajima}
   S. Tajima, T. Ido, S. Ishibashi, T. Itoh, H. Eisaki, Y. Mizuo, T. Arima, 
   H. Takagi, and S. Uchida, Phys. Rev. B {\bf 43}, 10496 (1991).
\bibitem{Thomas93}
   G. A. Thomas, D. H. Rapkine, S-W. Cheong, and L. F. Schneemeyer, Phys. Rev.
   B {\bf 47}, 11369 (1993).
\bibitem{Gervais}
   F. Gervais, P. Echegut, J. M. Bassat, and P. Odier, Phys. Rev. B {\bf 37}, 
   9364 (1988).
\end{references}
\end{document}